\documentclass[twoside]{article}
\usepackage{graphicx}
\usepackage{fancyhdr}
\usepackage{multicol}
\usepackage{styl-ap}

\begin{document}
\title{First hints of pressure waves in a helical extragalactic jet: S5~0836+710}
\author{Manel Perucho\work{1}}
\workplace{Departament d'Astronomia i Astrof\'{\i}sica. Universitat de Val\`encia. C/Dr. Moliner 50, 46100, Burjassot, Valencian Country, Spain.}
\mainauthor{manel.perucho@uv.es}
\maketitle

\begin{abstract}%
One of the open questions in extragalactic jet Astrophysics is related to the nature of the observed radio jet, namely whether it 
traces a pattern or the flow structure itself. In this paper I summarize the evidence collected for the presence of waves in extragalactic jets. 
The evidence points towards the peak of emission in helical jets corresponding to pressure-maxima of a wave that is generated within the core region and propagates downstream. Making use of a number of very long baseline interferometry (VLBI) observations of the radio jet in the quasar S5~0836+710 at different frequencies and epochs, Perucho et al. (2012a) were able to observe wave-like behavior within the observed radio-jet. The ridge-line of the emission in the jet coincides within the errors at all frequencies. Moreover, small differences between epochs at 15~GHz reveal wave-like motion of the ridge-line transversal to the jet propagation axis. The authors conclude that the helicity is a real, physical structure. I report here on those results and discuss this result in the light of new results recently announced by other authors that confirm the presence of waves in the close-by object BL Lac (Cohen et al., in preparation), and the possible magnetic or hydrodynamic nature.
\end{abstract}


\begin{multicols}{2}
\section{Introduction}
 The parsec-scale structure of jets in active galactic nuclei (AGN) is mainly observed in the radio band, using the 
VLBI technique. The nature and properties of the emitting region as related to 
the flow are still scarcely known. Jets can be interpreted as flows because the Larmor radius of particles is very small 
when compared to the spatial scales of the problem studied (Blandford \& Rees 1974). This interpretation has different implications:
Such systems, which are composed of magnetic fields and particles propagating along the jet channel are 
expected to host the growth of different hydrodynamical and/or magnetic instabilities. The instabilities have been claimed to be the cause of many of the structures observed (knots, bendings, helices; see e.g., reviews in Hardee 2006, 2011, Perucho 2012). The instabilities are waves triggered by any external or internal perturbation that grow in amplitude with distance, as they are advected with the flow. Thus, if instabilities do really grow in jets, we would expect to observe wave-like structures and motions in extragalactic jets. Perucho et al. (2012a) used observations of the jet in S5~0836+710 at different frequencies and epochs and showed that the 
ridge line of this jet behaves as expected if it is interpreted as a pressure wave.

The luminous quasar S5~0836+710 is at a redshift $z=2.16$ and hosts a powerful
radio jet extending up to kiloparsec scales (Hummel et al. 1992). At this redshift,
$1\,\rm{mas} \simeq 8.4\,\rm{pc}$ (see, e.g., MOJAVE database). VLBI monitoring of the
source showed kink structures (Krichbaum et al. 1990) and yielded estimates of the bulk Lorentz factor
$\gamma_\mathrm{j}=12$ and the viewing angle $\alpha_\mathrm{j}=3^\circ$ of the
flow at milliarcsecond scales (Otterbein et al. 1998). The jet was observed
at 1.6 and 5~GHz with VSOP (VLBI Space Observatory Program, a Japanese-led 
space VLBI mission), and oscillations of the ridge-line were also observed (Lobanov et al. 1998, 2006).
These oscillations were interpreted as Kelvin-Helmholtz (KH) instabilities (Lobanov et al. 2006, Perucho \& Lobanov 2006). The source has also been monitored by the MOJAVE program (Lister et al. 2009). Finally, it has been suggested that the lack of a collimated jet structure and hot-spot at the arc-second scales is due to jet disruption, as indicated by the growth in the amplitude of the helical structure (Perucho et al. 2012b).

\section{Evidence for waves in the jet of S5~0836+710}

 Here I summarize the evidence that points to the presence and development of waves in the jet of S5~0836+710, as given by Perucho et al. (2012a). The different observations of the jet at different frequencies ranging from 1.6~GHz to 43~GHz, for a given epoch, give the same ridge-line positions within the errors. It needs to be determined whether the small scale ripples in the ridge-line are caused by uv-coverage effects at the different epochs/frequencies, which is plausible. However, the large scale motion coincides for all frequencies and this can only happen if the observed structure corresponds to a physical one. In addition, high resolution images at 15~GHz confirm that the ridgr-line position does not necessarily coincide with the centre of emission of the radio jet, as expected from an asymmetric pressure maximum triggering a helical instability in the jet. Within the first milliarcsecond, high resolution images of the jet allowed the measurement of transversal displacements, which show a clear wave-like oscillation pattern with distance. The obtained ridge-line velocities at these scales are superluminal, but this is due, according to the authors, to: 1) relativistic wave velocities, 2) a small-scale oscillation of the core, very difficult to detect in this distant source, but observed in others (e.g., M~81, see Mart\'{\i}-Vidal et al. 2011), and 3) displacements of the ridge-line being smaller than the errors in the determination of the position. The amplitude of the oscillation grows along the jet propagation direction, which can be due to the coupling to an unstable mode of the KH or current-driven (CD) instabilities.
 
 
   The evidence collected in the previous paragraph led the authors to conclude that the ridge-line of the jet is tracing a pressure-maximum of a wave within the jet, and that even the radio-core is subject to the oscillation produced by the wave.  

   Recently, Cohen et al. (2013) presented new results of the nearby source BL~Lac (in the conference \emph{The innermost regions of relativistic jets and their magnetic fields}, held in Granada 10th-14th June, 2013), using long-term monitoring of the source at 15~GHz (Cohen et al. in preparation), which confirm that the ridge-line of emission within the jet shows a wave-like behavior. In this case, the authors claim that the oscillation of the ridge-line is caused by a strong magnetic field. In addition, Grossberger et al. (2012, and private communication) have followed the ejection and evolution of components in the broad-band radiogalaxy 3C~111 and shown that the position angles of the injected components vary with time and fill a wide channel after several years. This channel is wider than the observed jet at each epoch. This result also points to an oscillation of the bright regions of the jet (necessarily the same as that identified with the ridge-line in other works) within a wide channel in which the observed radio jet is embedded.

\section{Discussion}
  
  \subsection{On the nature of the waves}
  
 Given the evidence found for the presence of waves in jets, we can now ask which kind of waves these are and what is their origin. Regarding the latter, it is still unclear whether the waves show any regular periodical pattern or not, but the long-term oscillations could be attributed to periodical or quasi-periodical processes in the injection region, such as precession. Shorter-term oscillations can be attributed to perturbations triggered by asymmetries in the jet itself or in the ambient medium, lateral winds or entrainment of clouds that rotate around the galactic centre from a side of the jet, for instance. However, all these processes are extremely difficult to detect and isolate.
   
   As far as the nature of the waves, there is (already) open debate as to whether they are magnetic or purely hydrodynamical waves. The claim that the waves correspond to magnetic instabilities is based on the fact that the observed radio components seem to follow the ridge-line as they propagate (Cohen et al., in preparation): If the jet ridge line is identified with a kink instability in the magnetic field of a magnetically dominated jet, the flow would be forced to follow this kink. However, let us consider the possibility that the ridge-line corresponds to the pressure maximum produced by the kink (e.g., Mizuno et al. 2011). This implies that within a given jet cross-section we can find a high-pressure region and a low-pressure one. In this case, the passage of a ballistic component following a straight path along the jet, could enhance the emission of the whole jet, but the effect would be more visible on the ridge-line (high-pressure region). As a result, any radio-component fitted to the region would be centered there. 
         
    Perucho et al. (2006) showed, via numerical simulations, that an asymmetric perturbation at the base of a relativistic jet propagates downstream and naturally creates a pressure maximum helix, which can eventually distort the jet flow and force it into a helical path. In this view, injected components associated with perturbations in flow density at injection that produce shock waves, end up following the helical path when they have dissipated most of their kinetic energy. Previous to this, they propagate ballistically. 
    
    A long-standing cross-debate about jet physics is related to the ballistic nature of the flow in apparently helical radio jets. This point again introduces a difference in behavior, but still not a way to distinguish, between a kinetically dominated flow and a magnetically dominated one. If the jet is strongly magnetized, the flow will always follow the field lines and will thus have a twisted path if the field is helical, whereas in the kinetically dominated case, the flow could follow a straight stream-line even if the radio-jet shows a helical structure (because it corresponds to the bright high pressure region within a wide channel). This is true while the amplitude of the instability is not large, but when it grows to nonlinear values, the jet flow starts to deviate from its original trajectory by the pressure gradients within the jet (see, e.g., Nakamura \& Meier 2004, Mizuno et al. 2011, Perucho et al. 2005, 2006 for the different cases). Thus, a purely hydrodynamical jet could produce observational patterns similar to a strongly magnetized jet, and further work is needed to understand the physics behind the ridge-line dynamics.

    Moreover, even if no perturbations are injected in the jet, i.e., the flow is close to stationary, then the fitted components in the radio jet would all be associated with the ridge-line (see, e.g., Hardee 2003), just because it corresponds to the brightest regions within the jet. If the ridge-line corresponds to a propagating wave, then the fitted-components would also move along the jet and, of course, on top of the ridge-line. It is thus crucial to disentangle whether the components correspond to waves propagating along the jet and whether or not they follow the ridge-line for understanding its nature.

 \subsection{The case of S5~0836+710}  
   
      The wave associated with the ridge-line could couple to a growing instability, as indicated by the growth in amplitude of the helix with distance in the case of S5~0836+710. It is difficult to distinguish between current-driven (CD) instability and KH instability modes, both being solutions to the linearized relativistic, magnetized flow equations and both being possible sources of helical patterns. Recent work on CD instability by Mizuno et al. (2011) shows that a helical kink propagates with the jet flow if the velocity shear surface is outside the characteristic radius of the magnetic field, i.e., approximately the radius at which the toroidal magnetic field is a maximum. If the observed pattern corresponds to a CD kink instability, the observed transversal oscillation in the jet of 0836+710 requires that the kink be moving with the flow and implies that the transversal velocity profile is broader than the magnetic field profile, i.e., the velocity shear surface lies outside the characteristic radius of the magnetic field. Thus, in this case the jet would have a magnetized spine surrounded by a particle dominated outer region and a CD kink in the spine would move with the flow.. 
   
    The measurement of the opening angle reveals a very similar value at all frequencies for which we have significant measurements, which would favor the radio-jet tracing the whole channel, because the coincidence in the opening angles is less probable in the case of the radio jet corresponding to different regions across the jet. Thus, it seems that the radio emission is generated in the same region across the jet at all frequencies and that the jet flow indeed follows a helical path. As stated above, non-ballistic motion is compatible with a magnetically dominated flow, but also with the development of KH-instability modes to large amplitude, because they can dissipate kinetic energy into internal energy and decelerate the flow. It has been reported that the development of the helical KH instability can force the jet flow into a helical path, albeit different from the helicity of the pressure maximum, the latter showing larger amplitude (Hardee 2000). The larger the Lorentz factor of the flow and the shorter the wavelength of the mode, the more different is the helicity of the flow as compared to that of the pressure maximum. This can be understood in terms of the larger inertia of the jet flow with increasing Lorentz factor (Hardee 2000, Perucho et al. 2005). In summary, the fact that the flow is non-ballistic when the amplitude of the perturbation is large does not necessarily imply that the flow is magnetically dominated and that the waves are magneto-sonic waves.

\subsection{The core}
          One other important consequence of the work presented in Perucho et al. (2012a) is the possible small-scale oscillation of the radio-core around an equilibrium position. This effect has been reported by Mart\'{\i}-Vidal et al. (2011) for M~81, using the phase-referencing technique, and it could have important implications for the nature of the radio-core.  
 
 \section{Conclusions}
  
    It is possible to associate the ridge-line of emission of helical jets to wave patterns tracing a (thermal or magnetic) pressure maximum produced by the growth of instabilities. The core of the jet could well also be oscillating and this motion should be tested and corrected in order to study the parsec-scale kinematics of the radio-components in jets. Further combined observational and theoretical studies like this one are required to get more information on the nature of the growing instability and on the properties of the jet flow.



\thanks
I acknowledge financial support by the Spanish ``Ministerio de Ciencia e Innovaci\'on''
(MICINN) grants AYA2010-21322-C03-01 and AYA2010-21097-C03-01. 
I also acknowledge P.E. Hardee, J.M. Mart\'{\i}, M. Cohen, and D.L. Meier for interesting discussions on the topic.

\bigskip
\bigskip
\noindent {\bf DISCUSSION}

\bigskip
\noindent {\bf SERGIO COLAFRANCESCO:} Is there any radio polarization measurement that can be used to better describe the helical structure of the jet and, if so, what are the constraints that can be set on the acceleration mechanisms at the base of the jet?

\bigskip
\noindent {\bf MANEL PERUCHO:} Krichbaum et al. (1990) reported polarization measurements and claim that the field is aligned with the jet axis. If their interpretation is correct, it would be difficult to attach the helical morphology of the jet to the magnetic field. Nevertheless, new polarization measurements should be performed to check whether the magnetic field plays any crucial role in the generation of the helical structure. Regarding your second question, it is difficult to observe the acceleration region in this jet, but we have evidences of component acceleration in other AGN jets, included in the MOJAVE sample.

\bigskip
\noindent {\bf HERMAN MARSHALL:} If the perturbations are unstable and disrupt the jet at kpc scales, how do you explain that most jets seem to propagate to tens or hundreds of kpc?

\bigskip
\noindent {\bf MANEL PERUCHO:} In that sense, this is a unique object, because it is classified as an FRII in terms of its radio luminosity, but its morphology denies (see Perucho et al. 2012b). This is one of the reasons why we think that, in this sole case, the jet could be destroyed by the growth of the instability, unlike all other known FRII jets for which we can observe the kiloparsec scale structure. 

\end{multicols}
\end{document}